\documentclass[%
 aip,
 amsmath,amssymb,
reprint,%
floatfix,
]{revtex4-1}

\usepackage{graphicx}% Include figure files
\usepackage{dcolumn}% Align table columns on decimal point
\usepackage{comment}
\usepackage{bm}% bold math
\usepackage{multirow}
\usepackage{csquotes}
\usepackage{algpseudocode}
\usepackage[mathscr]{euscript}
\usepackage{amsmath}
\usepackage{bbm}
\usepackage{xcolor}

\newcommand\bra[1] {\langle {#1} |}
\newcommand\ket[1] {| {#1} \rangle}
\newcommand\braket[2] {\langle {#1} | {#2} \rangle}
\newcommand{\thezi}{Z_I}
\newcommand{\theb}{\mathbf{B}}
\newcommand{\theG}{\mathbf{G}}
\newcommand{\thennuc}{N_\mathrm{nuc}}
\newcommand{\thenel}{N_\mathrm{el}}
\def\elab{}
\def\ilab{\text{i}}

\begin{document}

\title{Unitary Coupled-Cluster for Quantum Computation of Molecular Properties in a Strong Magnetic Field
}

\author{Tanner Culpitt}
\email{tculpitt@wisc.edu}
\affiliation
{Theoretical Chemistry Institute and Department of Chemistry, 
University of Wisconsin-Madison, 1101 University Ave, Madison, Wisconsin 53706 USA}
\author{Erik I. Tellgren}
\affiliation
{Hylleraas Centre for Quantum Molecular Sciences,  Department of Chemistry, 
University of Oslo, P.O. Box 1033 Blindern, N-0315 Oslo, Norway}
\author{Fabijan Pavo\v{s}evi\'{c}}
\affiliation{Algorithmiq Ltd., Kanavakatu 3C, FI-00160 Helsinki, Finland}

%\date{\today}

\begin{abstract}
In truncated coupled-cluster (CC) theories, non-variational and/or generally complex ground-state energies can occur. This is due to the non-Hermitian nature of the similarity transformed Hamiltonian matrix in combination with CC truncation. For chemical problems that deal with real-valued Hamiltonian matrices, complex CC energies rarely occur. However, for complex-valued Hamiltonian matrices, such as those that arise in the presence of strong magnetic fields, complex CC energies can be regularly observed unless certain symmetry conditions are fulfilled. Therefore, in the presence of magnetic fields, it is desirable to pursue CC methods that are guaranteed to give upper-bound, real-valued energies. In this work, we present the first application of unitary CC (UCC) to chemical systems in a strong magnetic field. This is achieved utilizing the variational quantum eigensolver (VQE) algorithm applied to the unitary coupled-cluster singles and doubles (UCCSD) method. We benchmark the method on the H$_2$ molecule in a strong magnetic field, and then calculate UCCSD energies for the H$_4$ molecule as a function of both geometry and field angle. We show that while standard CCSD can yield generally complex energies that are not an upper-bound to the true energy, UCCSD always results in variational and real-valued energies. We also show that the imaginary components of the CCSD energy are largest in the strongly correlated region. Lastly, the UCCSD calculations capture a large percentage of the correlation energy.
 \end{abstract}

\maketitle

\section{Introduction}

Correlated methods that are cost-efficient are highly desirable in quantum chemistry. One popular option is the coupled-cluster (CC) method,~\cite{crawford2000introduction,bartlett2007coupled,shavitt2009many} which is equivalent to full configuration interaction (FCI) in the full CC (FCC) limit. A commonly used variant of CC is the coupled-cluster with singles and doubles (CCSD) method, which truncates the cluster operator at double excitations and thereby achieves computational feasibility while maintaining an adequate level of accuracy for a wide range of chemical systems. Truncation of the cluster operator at higher orders systematically improves the method's performance, albeit at an increased computational cost. Despite its success, it is worth noting that CCSD is not variational,~\cite{crawford2000introduction,bartlett2007coupled,shavitt2009many} and it is also not guaranteed to give real-valued energies.~\cite{thomas2021complex}

The CC methods have found application for the nonperturbative study of molecular properties in strong magnetic fields, including ground-state variants such as CCSD~\cite{Stopkowicz2015} and CCSDT~\cite{hampe2020full} as well as corresponding equation-of-motion-CC (EOM-CC) variants for excited states.~\cite{Hampe2017,Hampe2019,hampe2020full} The development of these CC methods is a part of ongoing work in the theoretical prediction of molecular electronic structure and dynamics in the presence of strong magnetic fields,~\cite{Lange2012,Tellgren2012,FURNESS_JCTC11_4169, Sun2019, Sen2019, Irons2021,Wibowo2021,David2021,Ceresoli2007,Culpitt2021,Peters2021,Monzel2022} and is especially applicable in the astrophysical domain for the identification of atoms and molecules in stellar objects such as magnetic white dwarf stars.~\cite{hollands2023dz}

In most quantum chemical contexts, where the Hamiltonian matrix is real-valued, the CCSD energy will also be real-valued. Exceptions occur near conical intersections, where complex CC energies can arise.~\cite{kjonstad2017resolving,thomas2021complex} However, in the presence of strong magnetic fields, the Hamiltonian matrix is generally complex.~\cite{Stopkowicz2015,Stopkowicz2018} Due to the non-Hermitian nature of the CCSD similarity transformed Hamiltonian, it has recently been reported that in a magnetic field complex CCSD energies can occur more frequently than in the zero-field case.~\cite{thomas2021complex} Indeed, complex energies may even be the norm unless certain symmetries are obeyed between the molecule and the magnetic field.~\cite{thomas2021complex} Due to the utility of the CC methods, it is desirable to explore variational alternatives to CCSD in the magnetic field paradigm that will not yield complex energies.

One approach is the unitary CC (UCC) method, which gives upper-bound and real-valued energies.~\cite{bartlett1989alternative,taube2006new} Another important advantage of the UCC method over the CC method is its good performance for the simulation of selected molecular processes where strong correlation is significant, such as covalent bond-breaking problems.~\cite{cooper2010benchmark} However, one drawback of UCC is the lack of a naturally terminating Baker-Campbell-Hausdorff (BCH) expansion for the similarity transformed Hamiltonian, which renders the resulting amplitude equations seemingly impossible to solve in principle due to an infinite number of terms.~\cite{taube2006new} In practice, however, it is possible to solve the UCC problem in a numerically exact way via Taylor expansion of the cluster operator on the reference wave function, and truncation of the series after contributions reach a negligibly small magnitude.~\cite{cooper2010benchmark,Evangelista_JCP2011,Kohn_JCP2022} This procedure can be performed in conjunction with direct minimization of the energy~\cite{Kohn_JCP2022} using the Wilcox identity,~\cite{Wilcox_JMP2004, VanVoorhis_JCP2001} or by solving projected amplitude equations.~\cite{Evangelista_JCP2011}

Although it is possible to solve UCC in a numerically exact way on a classical machine, the method scales exponentially even in case of truncated cluster operators. As an alternative to these schemes, one can also solve the truncated UCC problem using the variational quantum eigensolver (VQE) algorithm on quantum devices at a polynomial cost.~\cite{peruzzo2014variational} The VQE algorithm is a hybrid quantum-classical algorithm, where a quantum computer is used for preparation of the wave function and measurement of the energy, and a classical computer is used for parameter optimization.~\cite{peruzzo2014variational} Due to its hybrid form, it has become one of the most promising quantum algorithms for noisy intermediate-scale quantum (NISQ) devices.~\cite{preskill2018quantum} Over the past several years, the VQE algorithm was successfully implemented on various quantum computing architectures for calculation of ground-state energies of small molecules.~\cite{peruzzo2014variational,o2016scalable,shen2017quantum,hempel2018quantum,rossmannek2023quantum} Therefore, pursuing the development of UCCSD solved with the VQE for molecular systems in magnetic fields is highly desirable given its relevance to present and future quantum information technologies as well as its beneficial capability to yield real-valued, variational energies.  

This work is organized as follows. Section II contains theoretical background pertaining to UCCSD. Section III presents dissociation curves for H$_2$ generated using UCCSD and FCI in a strong magnetic field, as well as energy surfaces as a function of field angle and geometry for the H$_4$ molecule. Summary and future directions are given in Section~IV.

\section{Theory}

We consider a system of $\thennuc$ nuclei and $\thenel$ electrons. We use the notation $\thezi$ and $\mathbf{R}_I$ for the atom number and position of nucleus $I$, respectively. We use $\mathbf{r}^{\elab}$ and $\mathbf{p}^{\elab}$ for the position operator and canonical momentum operator of an electron, respectively. The vector potential of a uniform magnetic field $\mathbf B$ at position $\mathbf u$ is given by $\mathbf A(\mathbf u) = \frac{1}{2} \theb \times (\mathbf u-\theG)$, where $\theG$ is the gauge origin. 

\subsection{Second-quantized Hamiltonian in a uniform magnetic field}

The nonrelativistic Schr\"odinger--Pauli Hamiltonian within the Born-Oppenheimer approximation (description beyond the Born-Oppenheimer approximation can also be incorporated~\cite{pavosevic2020chemrev,pavosevic2021neouccsd,Culpitt2023_neo}) of a molecular system in a uniform magnetic field can be written in second-quantization as
\begin{align}
\hat{H} &= \sum_{pq}h_{pq} a_p^\dagger a_q + \frac{1}{2}\sum_{pqrs}g_{pqrs} a_p^\dagger a_q^\dagger a_s a_r . \label{neo_ham_sq}
\end{align}
The core-Hamiltonian matrix elements are given by (in atomic units)
\begin{align}
&h_{pq}=\bra{p}\hat{h}\ket{q} \nonumber \\
&=\int \!\mathrm  d\mathbf{x} \phi_{p}^{*}(\mathbf{x}) \Big ( \frac{1}{2}[\bm{\sigma}\cdot(\mathbf{p} + \mathbf{A}(\mathbf{r}))]^2 -\sum_{I=1}^{N_\text{nuc}}\frac{Z_I}{\vert \mathbf{r} - \mathbf{R}_I\vert}\Big ) \phi_{q}(\mathbf{x}) \label{hcore},
\end{align}
and the two-electron matrix elements are given by
\begin{align}
&g_{pqrs}=\braket{pq}{rs} \nonumber \\
&=\int \!\mathrm  d\mathbf{x}_1 \mathrm d\mathbf{x}_2\phi_{p}^{*}(\mathbf{x}_1)\phi_{q}^{*}(\mathbf{x}_2)r_{12}^{-1}\phi_{r}^{}(\mathbf{x}_1)\phi_{s}^{}(\mathbf{x}_2) \label{tpi},
\end{align}
In the above equations, $\mathbf{x}$ are collective space-spin coordinates, and in Eq.~\eqref{hcore}, $\bm{\sigma}$ is the vector of Pauli matrices
\begin{align}
\bm{\sigma}^x = \begin{pmatrix}
0 & 1 \\
1 & 0
\end{pmatrix}, \;
\bm{\sigma}^y = \begin{pmatrix}
0 & -\ilab \\
\ilab & 0
\end{pmatrix}, \;
\bm{\sigma}^z = \begin{pmatrix}
1 & 0 \\
0 & -1
\end{pmatrix}.
\end{align}

\subsection{Unitary coupled-cluster}
Starting from the Hartree-Fock (HF) reference state $\ket{0}$, the UCC wave function ansatz is written as 
\begin{align}
\ket{\Psi_{\text{UCC}}} &= \text{e}^{\hat{T}-\hat{T}^{\dagger}}\ket{0}
\label{wf_def_unicc},
\end{align}
where $\hat{T}$ is the cluster operator given by
\begin{align}
\hat{T}=\sum_n^{N_{\text{el}}} \hat{T}_n \ , \ \hat{T}_n = \frac{1}{({n!})^2} \sum_{ij...}\sum_{ab...} t_{ij...}^{ab...} a_a^{\dagger} a_i a_b^{\dagger} a_j ... . \label{T_op}
\end{align}
The subscript $n$ in Eq.~\eqref{T_op} refers to the number of particle-hole pairs and indices $i,j,k,$... refer to occupied orbitals while indices $a, b, c,$... refer to virtual orbitals. The UCC energy is then found via minimization of the energy functional with respect to the generally complex cluster amplitudes  $t_{ij...}^{ab...}$ according to
\begin{align}
E_{\text{UCC}}=\underset{t}{\text{min}}\bra{0}\text{e}^{\hat{T}^{\dagger}-\hat{T}}\hat{H}\text{e}^{\hat{T}-\hat{T}^{\dagger}} \ket{0}. \label{E_UCC}
\end{align}
Truncation of the cluster operator at doubles ($\hat{T} = \hat{T}_1 + \hat{T}_2)$ yields the UCCSD approximation.

\section{Results}

In this section, we present potential energy surfaces (PESs) for the H$_2$ and H$_4$ molecules using UCCSD solved with the VQE algorithm. In the case of H$_2$, dissociation curves were generated for UCCSD and conventional FCI in order to benchmark the UCCSD results in a strong magnetic field (in this instance both methods are exact within a given basis). We then applied UCCSD to H$_4$. Because the electronic energy of molecules in a magnetic field depends parametrically on both the nuclear coordinates (here assuming the Born-Oppenheimer approximation) and the magnetic field vector, we have studied the PES of H$_4$ as a function of both of these variables. Comparisons were made to FCI and conventional CCSD. 

We have used the Jordan-Wigner transformation~\cite{jordan1993algebraic} to map the fermionic operators in the Hamiltonian and cluster operators $\hat{T}$ to the qubit basis which is implemented in OpenFermion.~\cite{mcclean2020openfermion} The Jordan-Wigner transform maps a fermionic creation operator to the qubit basis according to $a_p^\dagger = \bm{\sigma}_p^{-} \otimes^{q<p} \bm{\sigma}_q^z$, where $\bm{\sigma}_p^{\pm} = 1/2 (\bm{\sigma}_p^x \pm i\bm{\sigma}_p^y)$, and an annihilation operator is mapped analogously using $\bm{\sigma}_p^{+}$. The one-particle and two-particle integrals are provided via an interface to the LONDON program.~\cite{LondonProgram} Note that in a magnetic field, the Hamiltonian matrix and wave function will be generally complex. Additionally, there is a gauge-origin dependence in the Hamiltonian introduced by the magnetic vector potential. The LONDON software program addresses these complications, and is capable of nonperturbative, gauge-origin invariant calculations of molecular properties in strong magnetic fields using London orbitals (also known as gauge-including atomic orbitals (GIAOs)).\cite{London1937,Hameka1958,Ditchfield1976,Helgaker1991,Tellgren2008,Tellgren2012,Irons2017,Pausch2020} The basis set used for all calculations is a London orbital variant of the STO-3G basis, denoted L-STO-3G.

Variational optimization of the energy defined is Eq.~\ref{E_UCC} was performed using the Broyden-Fletcher-Goldfarb-Shannon (BFGS) algorithm implemented in SciPy.~\cite{virtanen2020scipy} Unlike the zero-field case, here the minimization problem is necessarily over complex cluster amplitudes. However, by treating the real and imaginary parts of the cluster amplitudes as separate variational parameters, the problem can be recast as a real optimization problem of double dimension.

\subsection{H$_2$ Molecule}

\begin{figure}[h]
\centering
\begin{tabular}{ll}
(a) \\
\includegraphics[width=0.48\textwidth]{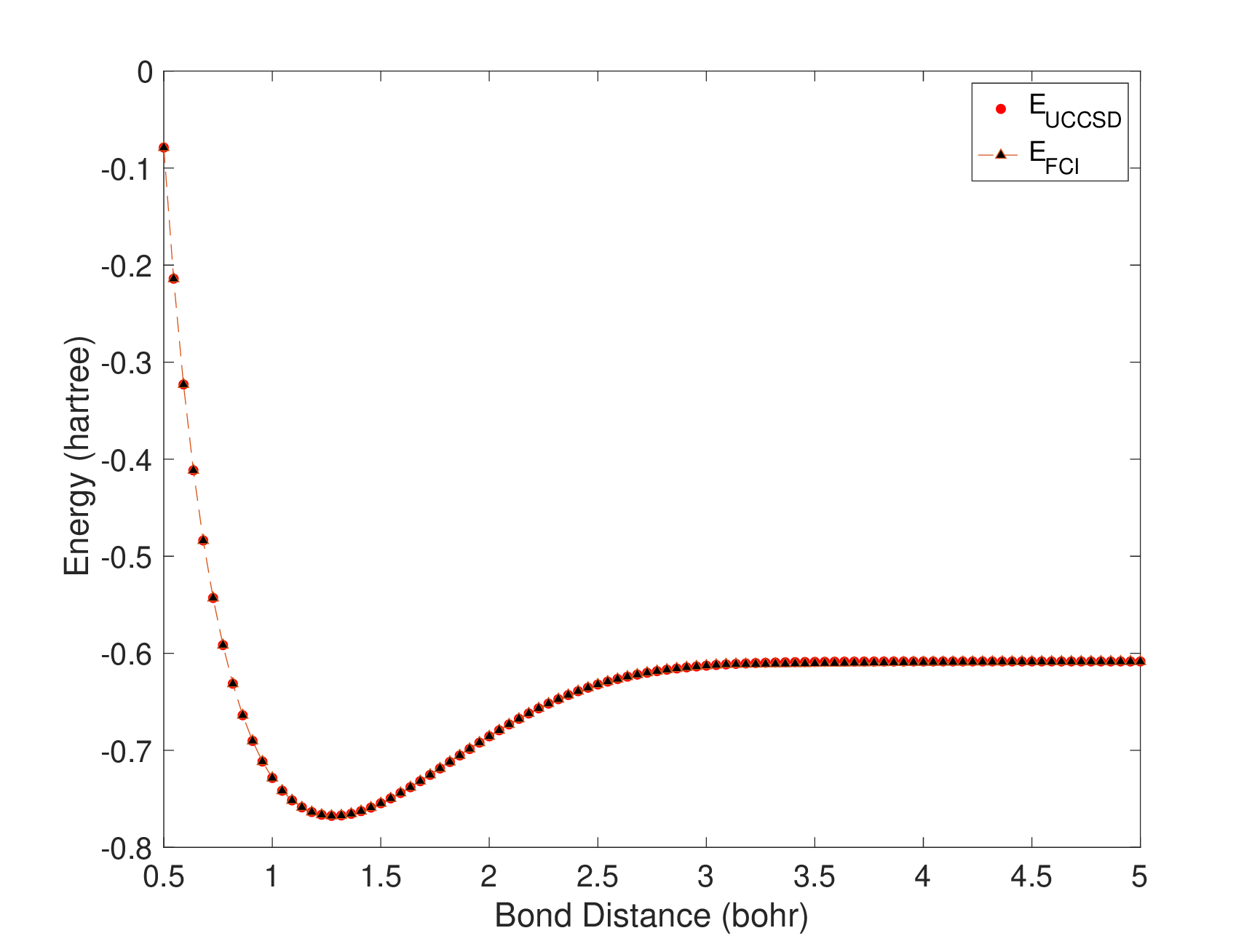} \\
(b) \\
\includegraphics[width=0.48\textwidth]{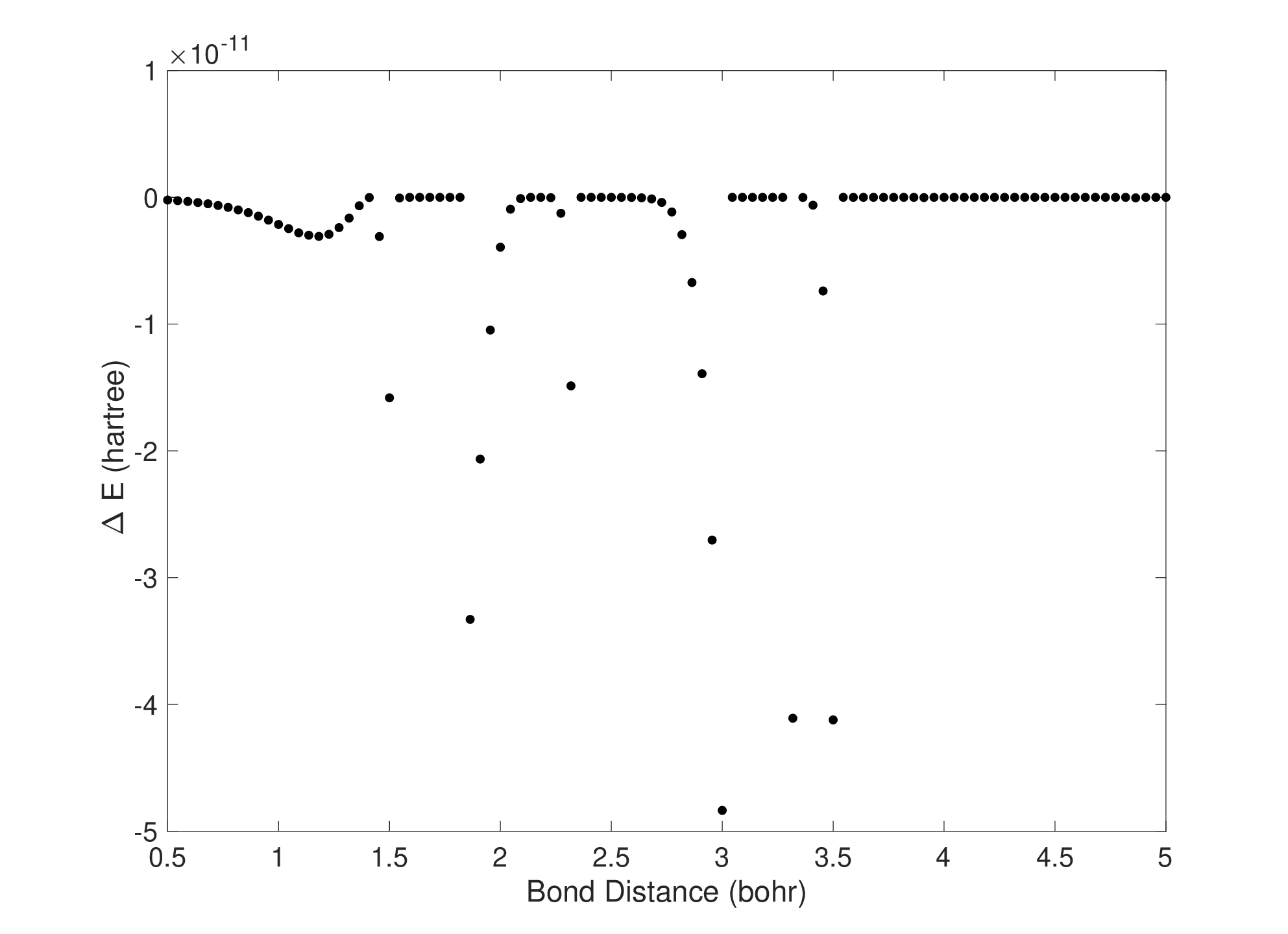} \\
\end{tabular}
\caption{(a) Dissociation curve for the H$_2$ molecule for the singlet state computed with UCCSD and FCI in a strong magnetic field of $|\mathbf{B}|=1.0 \ \text{B}_0$ oriented perpendicular to the molecule. (b) Difference between the FCI energy and  UCCSD energy. }
\label{figure01}
\end{figure}

We study this system primarily as a benchmark for the UCCSD implementation in strong magnetic fields. Dissociation curves for the singlet state were generated for the H$_2$ molecule using UCCSD solved via the VQE and conventional FCI. The UCCSD curve is plotted in Fig.~\ref{figure01}(a) at a magnetic field strength of 1.0B$_0$ (B$_0 \approx 235000$ T) oriented perpendicular to the molecular axis. The FCI results are identical to the UCCSD results as observed by the overlapping dissociation curves in Fig.~\ref{figure01}(a) and by the negligible difference between the FCI and UCCSD energies (magnitude on the order of $10^{-11}$ at maximal values) as shown in Fig.~\ref{figure01}(b).

\subsection{H$_4$ Molecule}

\begin{figure}[h]
\centering
\begin{tabular}{ll}
(a) \\
\includegraphics[width=0.48\textwidth]{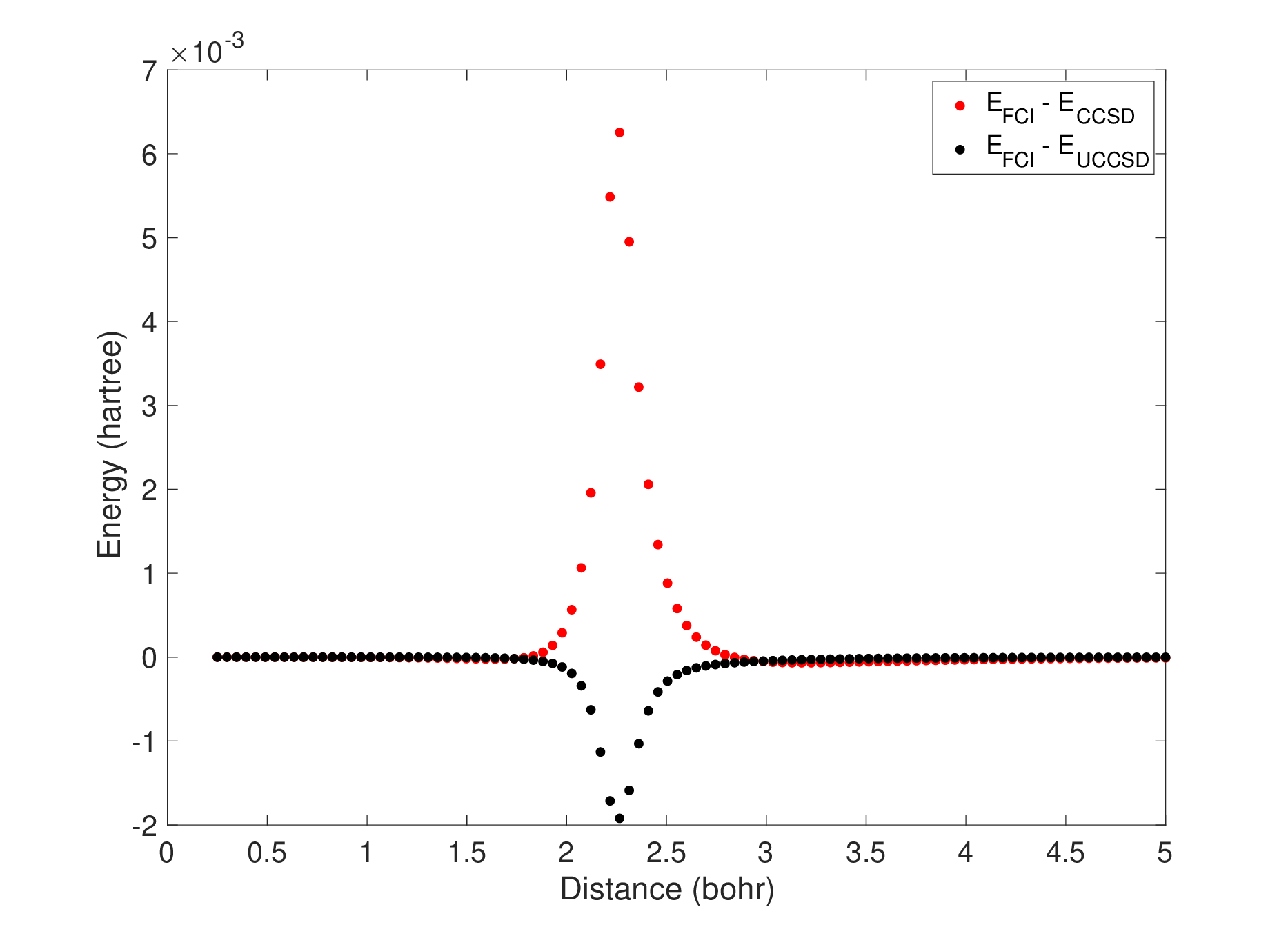} \\
(b) \\
\includegraphics[width=0.48\textwidth]{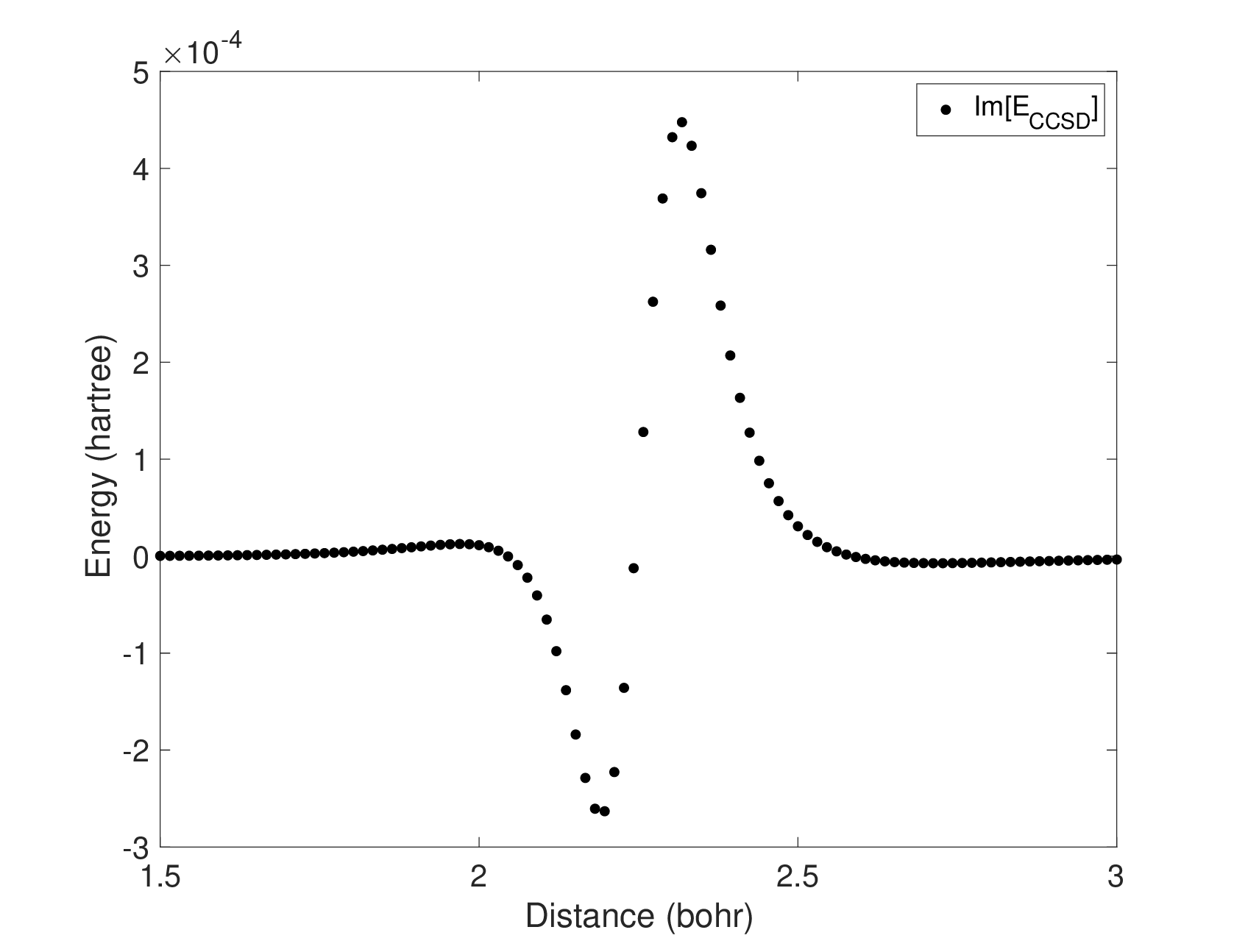} \\
\end{tabular}
\caption{ (a) Difference between the FCI energy and Re[CCSD] energy (red) and UCCSD energy (blue) of the H$_4$ molecule in a uniform magnetic field with components B$_x= 0.218$ B$_0$, B$_y=0.018$ B$_0$, B$_z=0.335$ B$_0$ as a function of geometry. Energies are plotted as a function of the distance $\gamma$ for the atomic coordinates H$_1$ = (0,0,0), H$_2$ = (0,0,2.32), H$_3$ = ($\gamma$,0,0), H$_4$ = ($\gamma$,0,2.32). (b) Imaginary component of the CCSD energy as a function of the H$_4$ distortion described above. Note the different x-axis limits for panels (a) and (b). }
\label{h4_geom_stretch}
\end{figure}

\begin{figure}[h]
\centering
\begin{tabular}{ll}
(a) \\
\includegraphics[width=0.48\textwidth]{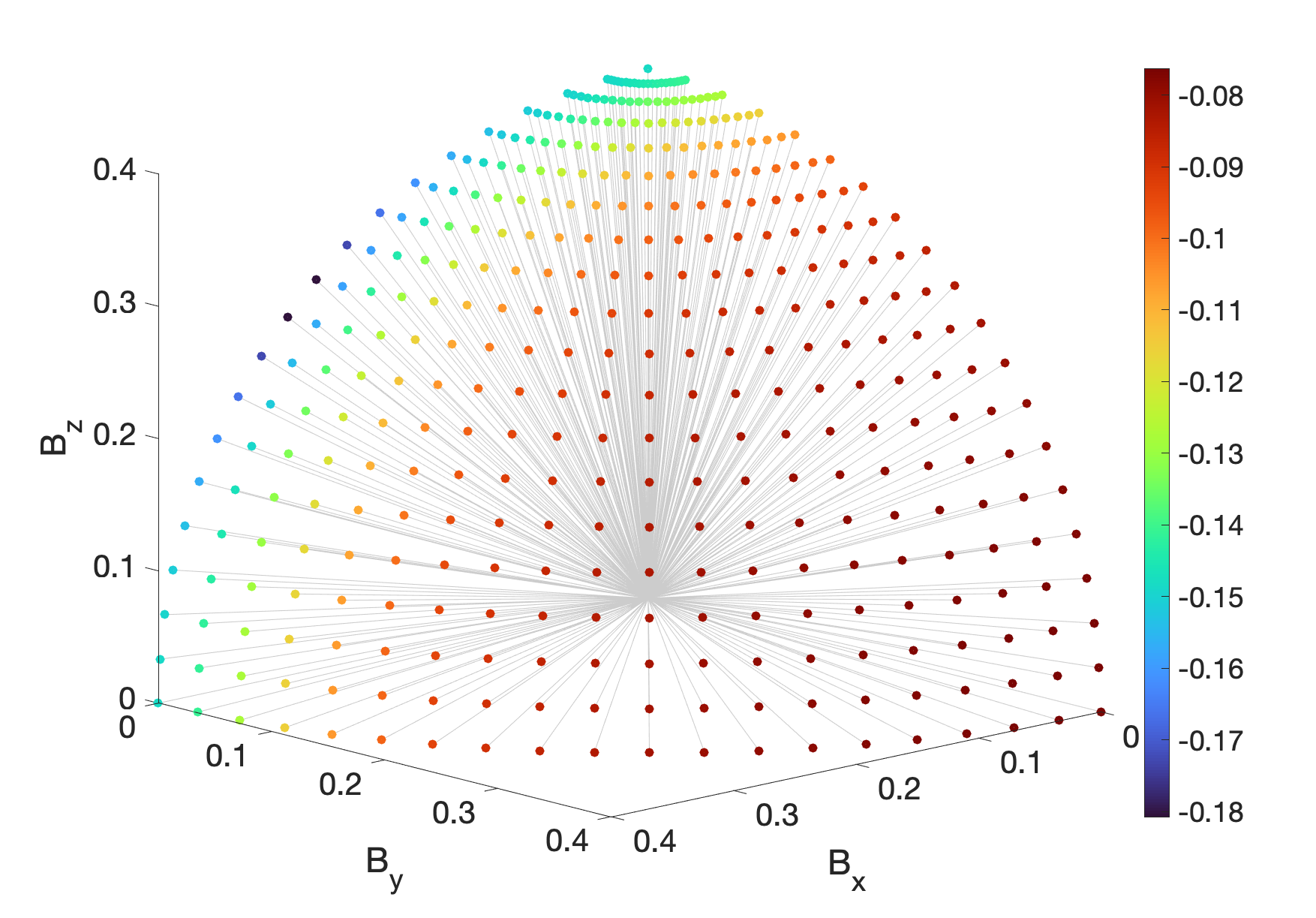} \\
(b) \\
\includegraphics[width=0.48\textwidth]{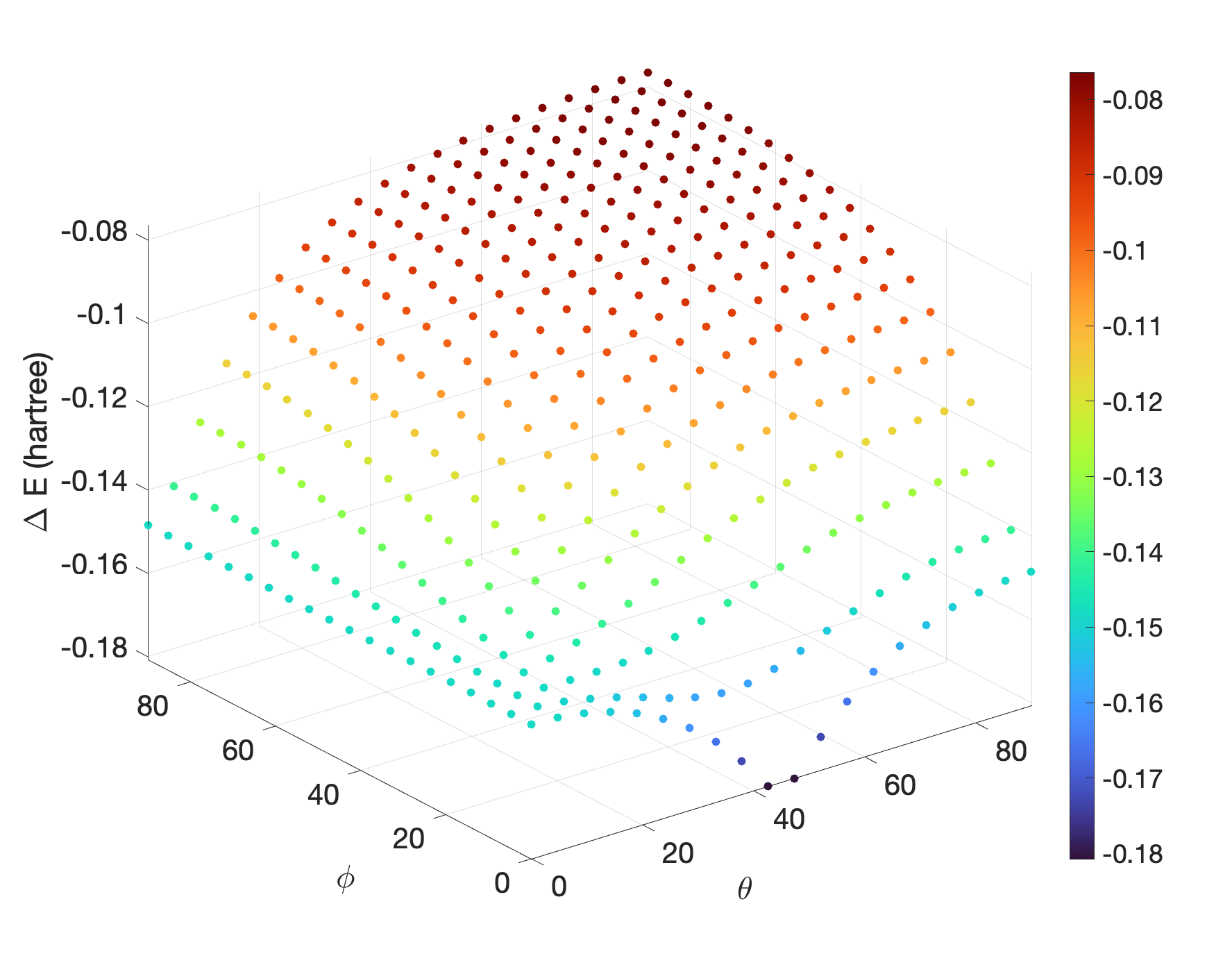} \\
\end{tabular}
\caption{ (a) Magnetic field vectors of magnitude 0.4 B$_0$ used to generate the FCI, CCSD, and UCCSD PESs as a function of field angle for the fixed H$_4$ geometry reported in the text. The color of $\mathbf{B}$ vector vertices matches the color of the corresponding point on the correlation energy plot given in the lower panel, i.e., the color bars of both panels are identical and reflect the value of the correlation energy (b) Correlation energy computed as a function of magnetic field angle for the fixed H$_4$ geometry reported in the text.}
\label{figure03}
\end{figure}

All H$_4$ PESs for FCI and UCCSD are for the singlet state, and the CCSD surfaces are generated beginning from an RHF reference. The energy of the H$_4$ molecule was calculated as a function of geometry for a  magnetic field of magnitude 0.4 B$_0$ with components B$_x= 0.218$ B$_0$, B$_y=0.018$ B$_0$, B$_z=0.335$ B$_0$. This magnetic field angle was chosen because it was observed to give a large imaginary component of the CCSD energy in the vicinity of the square geometric H$_4$ configuration. Rectangular distortion was performed horizontally in the x-z plane according to the displacement $\gamma$ for the atomic coordinates (given in bohr) H$_1$ = (0,0,0), H$_2$ = (0,0,2.32), H$_3$ = ($\gamma$,0,0), H$_4$ = ($\gamma$,0,2.32). The difference between the conventional FCI energy and both CCSD and UCCSD energies is plotted as a function of the distance $\gamma$ in Fig.~\ref{h4_geom_stretch}. Note that there is a region of strong correlation that ranges from $\approx 1.75$ -- 3 bohr, where both differences $E_\text{FCI} - \text{Re}[E_{\text{CCSD}}]$ and $E_\text{FCI} - E_{\text{UCCSD}}$ become large. However, it is observed that the UCCSD energy is always above the FCI energy in this region, while the CCSD energy is always below the FCI energy, which is not physical, and displays the non-variational character of CCSD. Additionally, imaginary components of the CCSD energy are largest in the strongly correlated region, being on the order of $10^{-4}$ hartree.
Next, the PES of the H$_4$ molecule was studied as a function of field angle for a fixed magnetic field strength of B = 0.4 B$_0$. We take inspiration from Ref.~\onlinecite{thomas2021complex}, where a similar procedure was carried out but for the water molecule. We use a square geometry for H$_4$, which is in the strongly correlated regime. The molecule is placed in the x-z plane, with coordinates (given in bohr) H$_1$ = (0,0,0), H$_2$ = (0,0,2.32), H$_3$ = (2.32,0,0), H$_4$ = (2.32,0,2.32). The magnetic field vector as a function of angle is given by
\begin{align}
\mathbf{B}(\theta,\phi)= 0.4\begin{pmatrix}
\text{sin}(\theta)\text{cos}(\phi) \\
\text{sin}(\theta)\text{sin}(\phi) \\
\text{cos}(\theta)
\end{pmatrix}
\end{align}
where $\theta$ is the angle from the positive z-axis and $\phi$ is the angle from the positive x-axis. Both $\theta$ and $\phi$ vary between $0^\circ$ and $90^\circ$  such that all vectors lie in the first octant (see Fig.~\ref{figure03}(a)). Three methods are employed: UCCSD with the VQE, conventional FCI, and conventional CCSD. The sampled magnetic field vectors are plotted in Fig.~\ref{figure03}(a) while the correlation energy is plotted in Fig.~\ref{figure03}(b). The vertices of the magnetic field vectors are colored according to the correlation energy to ease interpretation of the results. It is observed that the magnitude of the correlation energy tends to be largest for angles of $\theta$ and $\phi$ that are generally small, and diminishes as these angles grow mutually larger. Note that the correlation energy has a qualitatively inverse relationship with the absolute UCCSD energy plotted in Fig.~\ref{figure05}(a); that is, the correlation energy is lowest in regions where the absolute energy is highest and vice versa.

\begin{figure}[h]
\centering
\begin{tabular}{ll}
(a) \\
\includegraphics[width=0.48\textwidth]{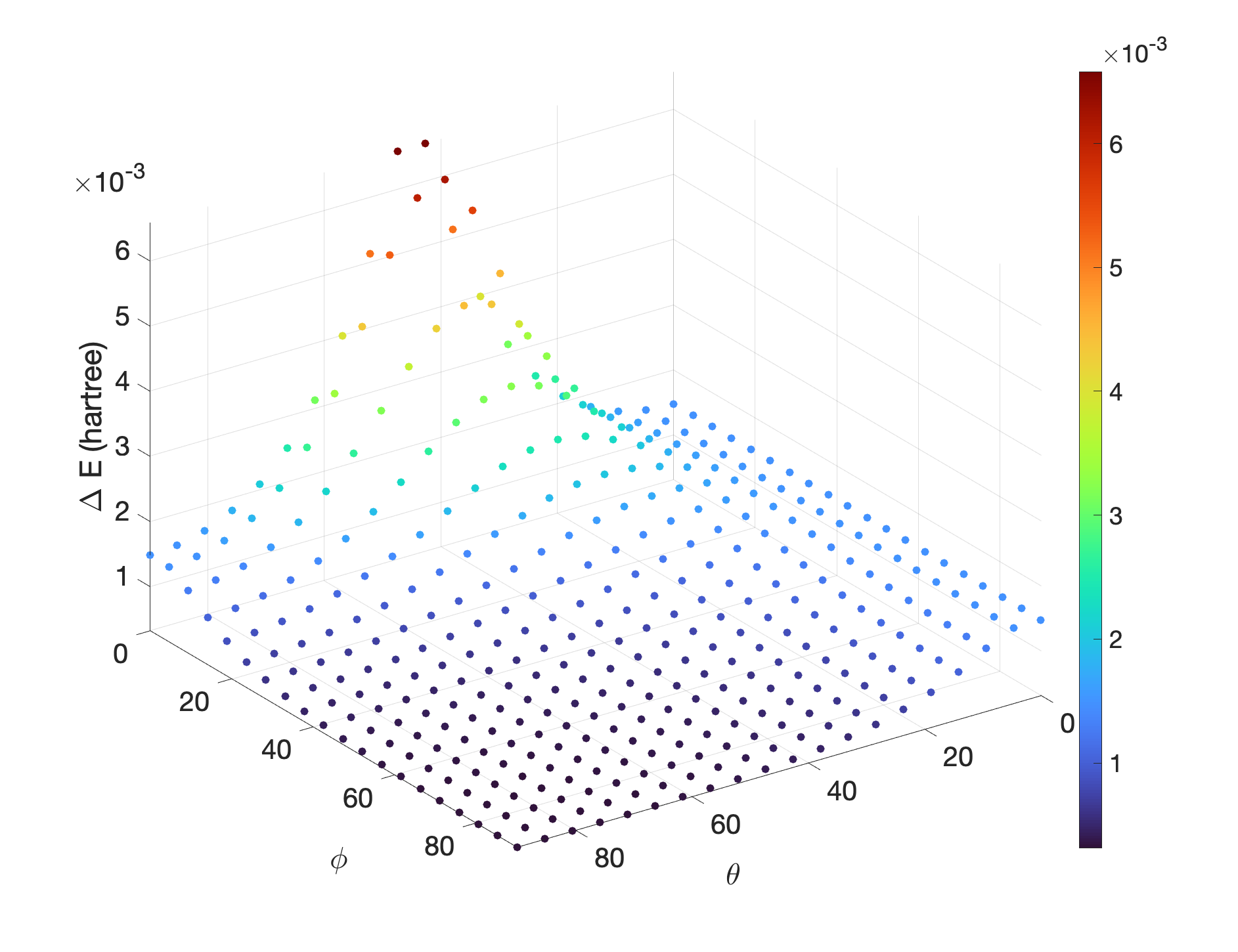} \\
(b) \\
\includegraphics[width=0.48\textwidth]{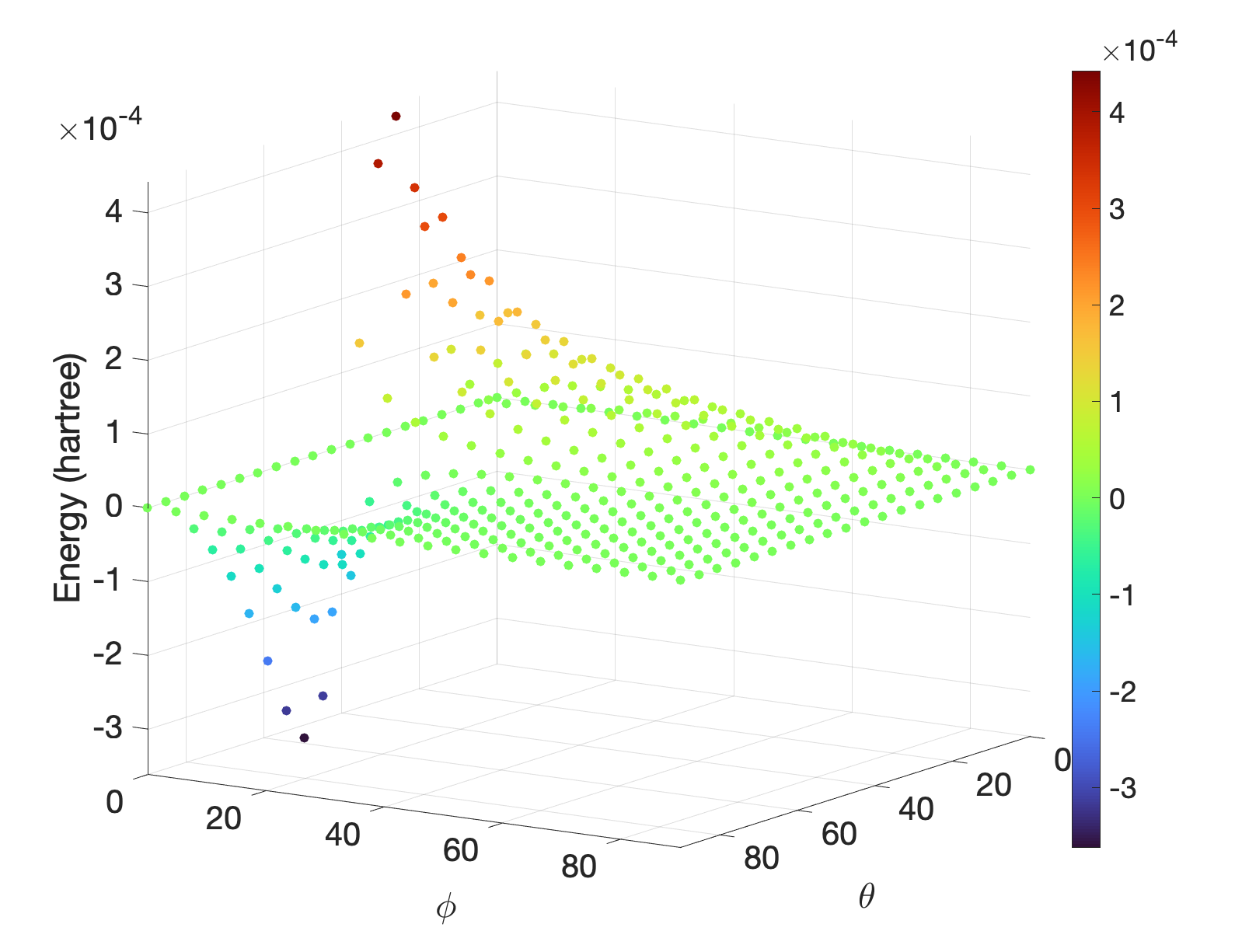} \\
\end{tabular}
\caption{ (a) Difference between the FCI energy and the real component of the CCSD energy as a function of magnetic field angle with $|\mathbf{B}|=0.4 \ \text{B}_0$ for the square geometry of the H$_4$ molecule as given in the main text (b) Imaginary component of the CCSD energy as a function of magnetic field angle for the same conditions as stated above. }
\label{figure04}
\end{figure}

\begin{figure}[h]
\centering
\begin{tabular}{ll}
(a) \\
\includegraphics[width=0.48\textwidth]{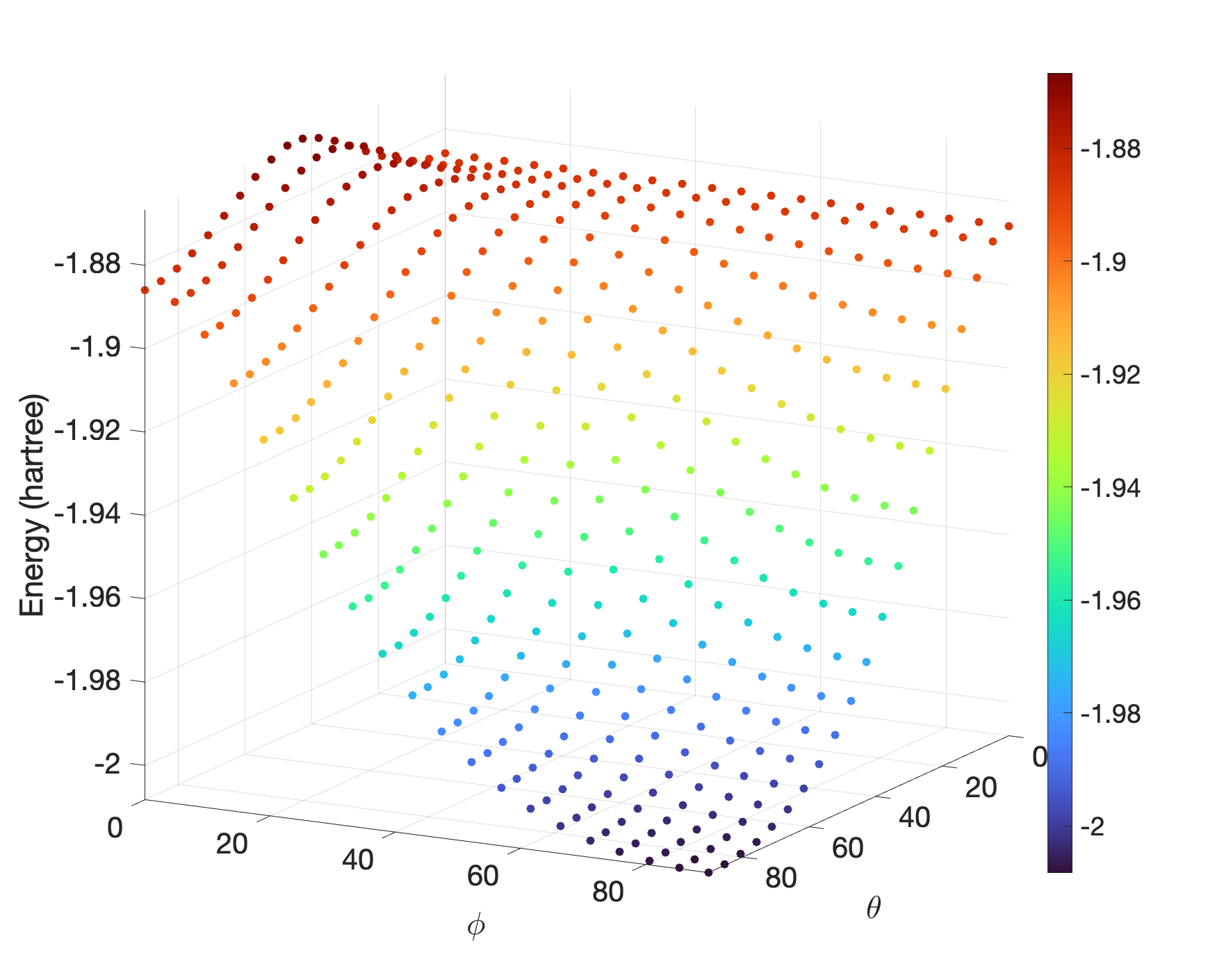} \\
(b) \\
\includegraphics[width=0.48\textwidth]{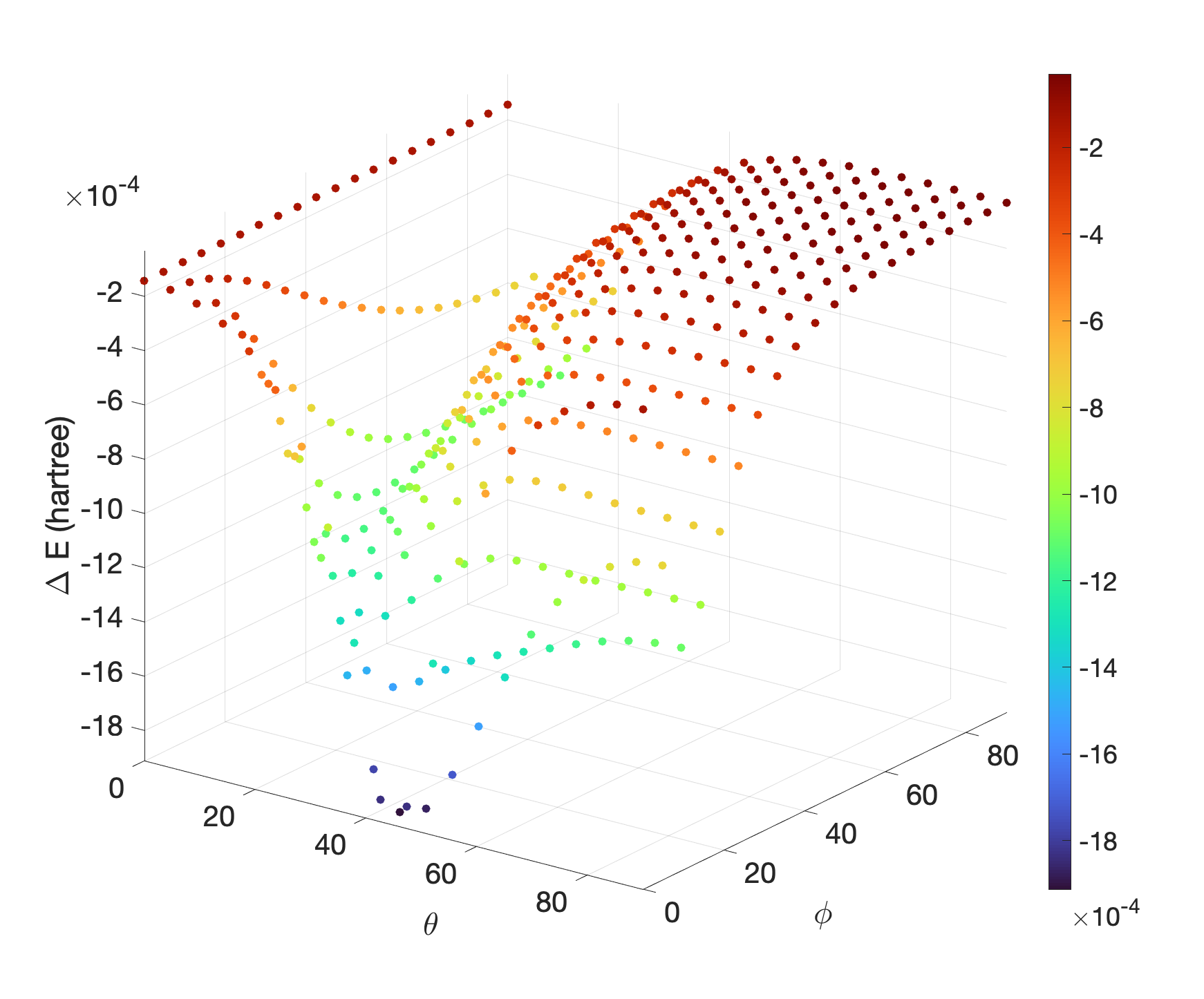} \\
\end{tabular}
\caption{(a) UCCSD energy computed with the VQE as a function of magnetic field angle with $|\mathbf{B}|=0.4 \ \text{B}_0$ for the square geometry of the H$_4$ molecule as given in the main text (b) difference between the FCI energy and the UCCSD energy as a function of magnetic field angle with $|\mathbf{B}|=0.4 \ \text{B}_0$ for the square geometry of the H$_4$ molecule as given in the main text.}
\label{figure05}
\end{figure}

The CCSD energies are found to be generally complex as a function of field angle, with the imaginary component being plotted in Fig.~\ref{figure04}(b). The magnitude of the imaginary part of the energy tends to be largest in regions where the magnitude of the correlation energy is largest. The largest magnitude of the imaginary part has a value of $4\times10^{-4}$ hartree and occurs at $\theta \approx 33^\circ$ and $\phi \approx 5^\circ$. Note that this is roughly 100 times larger than the highest absolute value of the imaginary component of the CCSD energy reported in Ref.~\citenum{thomas2021complex} for the water molecule. The magnitude of the correlation energy at this field angle is $1.6\times10^{-1}$ hartree, or about 400 times larger than the corresponding imaginary component. This ratio is roughly 200 times smaller than the analogous case for water reported in Ref.~\citenum{thomas2021complex}.

We find that the CCSD energies are real-valued when at least one component of the magnetic field vector vanishes, which is the case for the perimeter of the octant traced by the sampled magnetic field vectors. This observation is in agreement with Ref.~\citenum{thomas2021complex}, where similar results were obtained for the water molecule. However, we found that the real component of the CCSD energy (whether the total energy is purely real or generally complex) is everywhere lower than the FCI energy. This is readily observed in Fig.~\ref{figure04}(a), where the plotted difference $E_{\text{FCI}}-\text{Re}[E_{\text{CCSD}}]$ is everywhere positive. At large angles of $\theta$ and $\phi$ the CCSD energy approaches the FCI energy from below, but still remains lower. This is, of course, possible, given that CCSD is not variational. The lowest real component of the CCSD energy is $\approx 6\times 10^{-3}$ hartree below the FCI energy, and occurs at $\theta \approx 47^\circ$ and $\phi = 0^\circ$.

The UCCSD method is found to be a dramatic improvement over CCSD. First, all calculated energies are purely real-valued, which is already an improvement relative to CCSD. Second, because UCCSD is variational, we see in Fig.~\ref{figure05}(b) that the difference between the FCI energy and the UCCSD energy is everywhere negative, and also quite small, with the UCCSD energy being on the order of $10^{-4}$ to $10^{-3}$  hartree above the FCI energy. Even in the most strongly correlated regions, UCCSD recovers as much as $98.8 \%$ of the correlation energy. By contrast, it is difficult to even report the analogous quantities for CCSD in this case, because the CCSD energy is in most regions complex, and where it is real-valued, it is still lower than the FCI energy, which is not physical. We therefore conclude that UCCSD is a desirable alternative to CCSD in a strong magnetic field due to its accuracy and capability to yield variational, real-valued energies.

\section{Conclusions}

In this work, we have presented the application of the UCCSD method to molecular systems in a strong magnetic field using the VQE algorithm. The UCCSD method was benchmarked on the H$_2$ molecule and found to exactly reproduce the conventional FCI results. It was then applied to the H$_4$ molecule and compared against conventional FCI and CCSD. In a magnetic field, it is possible for the CCSD energy to be complex. This phenomena was observed when computing the CCSD energy at various geometries for a fixed magnetic field angle, and at various angles to a strong magnetic field for a fixed geometry. Moreover, the real part of the CCSD energy was found to be lower than the FCI energy in regions of strong correlation. Thus, we have investigated an example where the CCSD method fails on two fronts: (1) many field angles/geometries result in a complex energy (2) those energies that are real-valued may be below the FCI energy and therefore not physical.

By contrast, the UCCSD energies were found to be everywhere real-valued and an upper-bound to the FCI energy. Additionally, UCCSD recovered a significant portion of the correlation energy, even in regions where the magnitude of the correlation energy was large. Therefore, we believe that this approach is promising for future applications to systems in a strong magnetic field or in other physical situations where time-reversal symmetry breaking occurs. Future works include simulations of molecules in strong magnetic fields on real quantum devices which would require development of hardware-efficient techniques, such as ADAPT-VQE.~\cite{grimsley2019adaptive,tang2021qubit} Extensions of the developed methods to allow for calculations of excitation energies and excited state properties in magnetic fields is also of interest.~\cite{ollitrault2020quantum,pavosevic2023spinflip}

\section*{Acknowledgements}

We thank Francesco Evangelista for helpful discussions. This work was supported by the Research Council of Norway through ‘‘Magnetic Chemistry’’ Grant No.\,287950 and CoE Hylleraas Centre for Quantum Molecular Sciences Grant No.\,262695. The work also received support from the UNINETT Sigma2, the National Infrastructure for High Performance Computing and Data Storage, through a grant of computer time (Grant No.\,NN4654K).

\section*{Data Availability}

The data that support the findings of this study are available within the article.

\end{document}